\documentclass{article}
\usepackage{spconf,amsmath}
\usepackage{multirow}
\usepackage{url}
\usepackage[pdftex]{graphicx}

\title{Peer Collaborative Learning for Polyphonic Sound Event Detection}
\name{Hayato Endo$^{1}$ and ~Hiromitsu Nishizaki$^{1}$}
\address{$^{1}$ Integrated Graduate School of Medicine, Engineering, and Agricultural Sciences, \\ University of Yamanashi, Japan}
\begin{document}
\ninept
\baselineskip 10.6pt

\maketitle
\begin{abstract}
  This paper describes that semi-supervised learning called peer collaborative learning (PCL) can be applied to the polyphonic sound event detection (PSED) task, which is one of the tasks in the Detection and Classification of Acoustic Scenes and Events (DCASE) challenge. Many deep learning models have been studied to find out what kind of sound events occur where and for how long in a given audio clip. The characteristic of PCL used in this paper is the combination of ensemble-based knowledge distillation into sub-networks and student-teacher model-based knowledge distillation, which can train a robust PSED model from a small amount of strongly labeled data, weakly labeled data, and a large amount of unlabeled data. We evaluated the proposed PCL model using the DCASE 2019 Task 4 datasets and achieved an F1-score improvement of about 10\% compared to the baseline model. 
\end{abstract}

%
%
\section{Introduction}
\label{sec:intro}

In the present, speech recognition tasks have occupied a leading position in acoustic signal recognition, but it is unavoidable that various environmental sound processing, including speech, will become the acoustic information processing of the next era. Recently, the Detection and Classification of Acoustic Scenes and Events (DCASE) community \cite{DCASE} has been established, where various research tasks (challenges) related to environmental sound processing have been proposed, and researchers around the world are studying them.

In this paper, we focus on the research of ``Task 4: sound event detection in domestic environments'' \cite{Turpault2019} proposed in DCASE 2019 \cite{DCASE2019Workshop} and DCASE 2020 \cite{DCASE2020Workshop}, which is the task of identifying where a certain sound is occurring in an audio clip. When we focus on a certain moment in an audio clip, we have to consider all the sound events that may be occurring at the same time. This kind of task is called polyphonic sound event detection (PSED) \cite{Mesaros2016_MDPI}. Unlike monophonic sound event detection (MSED), which assumes that only one type of sound event occurs at a given moment, the PSED task is more difficult from various perspectives. For example, from a technical point of view, we have to estimate multiple sound types simultaneously, which is more difficult than MSED. The DCASE Task 4 is one of the PSED tasks.

In order to solve this PSED task using machine learning such as deep learning, a large amount of supervised labeled data is required to train a machine learning model. In this case, the supervised label is the data with the exact sound event tag of what kind of sound event occurs in which section of a sound file. This kind of supervised data is called the strongly labeled data. It would be easy to train the model if there is a large amount of such strongly labeled data. However, it takes much time and human resources to prepare the strongly labeled data. On the other hand, a labeling method only assigns a supervised label to an acoustic file with the information of what kind of sound event is contained in the file instead of labeling the exact interval of sound event occurrence. This type of supervised data is called weakly labeled data. Although the weakly labeled data is easier to prepare than the strongly labeled data, there is no doubt that labeling a large number of acoustic files is still expensive.

\begin{figure}[tb]
  \centering
  \includegraphics[width=0.99\columnwidth]{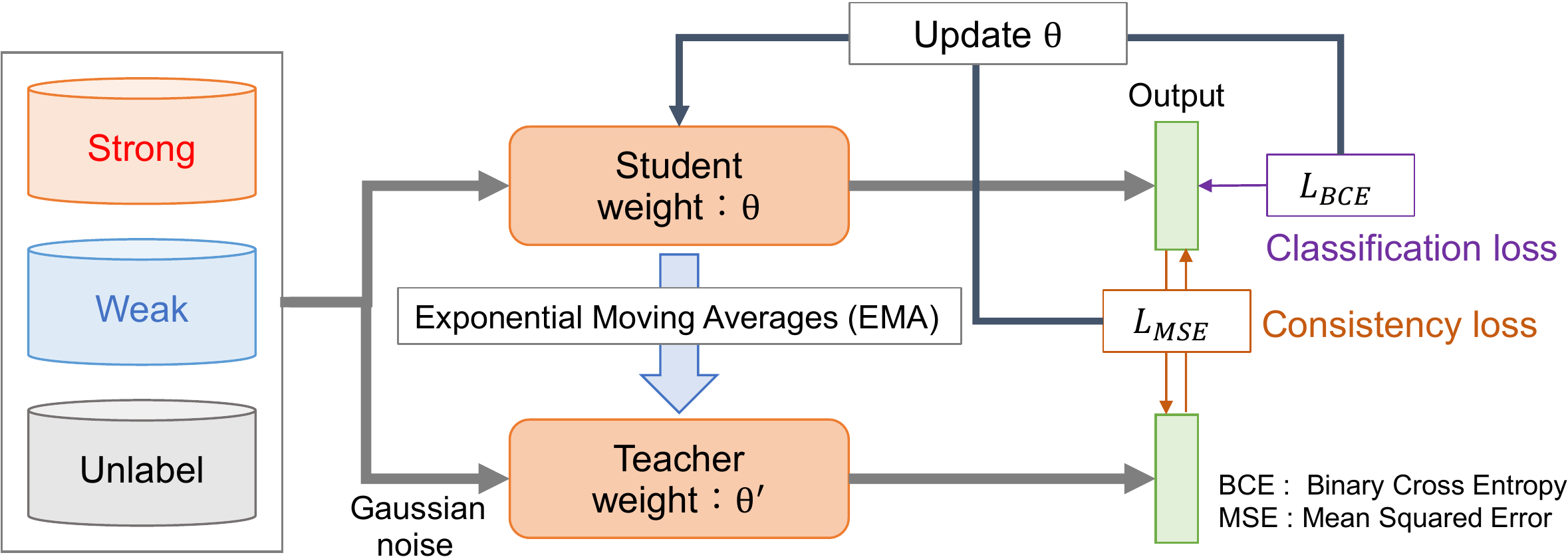}
  \vspace*{-2mm}
  \caption{A framework for a mean-teacher model.}
  \label{fig01}
\end{figure}

The DCASE 2019 and 2020 Task 4 provided a small amount of strongly labeled, weakly labeled, and a large amount of unlabeled data. Semi-supervised learning methods using these data are being actively studied. The baseline method provided by the Task 4 organizers also adopts semi-supervised learning, as shown in Figure \ref{fig01}. The baseline model is trained in the framework of a mean-teacher method \cite{Turpault2019} based on a student-teacher model \cite{44873, 8639635}. The model parameters based on the loss value are updated only in the student model, and the parameters of the student model are also reflected in the teacher model. In other words, the teacher model will accumulate the knowledge of the student model. By constraining the output of the teacher model to be the same as the output of the student model, the training of the student model is stabilized.

In this other previous work, for example, Lin et al. \cite{Lin2019} proposed a guided learning method to train the student-teacher model more efficiently. Park et al. \cite{Park2019b} suggested a tri-training approach, in which two different classifiers are used to acquire pseudo-labels from weakly labeled and unlabeled datasets and use them. Similarly, Ebbers et al. also proposed \cite{Ebbers2020} a method to train a classifier after assigning pseudo-labels to weakly labeled and unlabeled data sets. In addition, Kim et al. \cite{9312148} improved the mean-teacher method and proposed a two-stage distillation method using a fine-tuning model based on semi-supervised loss, which achieved state-of-the-art performance on the DCASE 2020 Task 4 test set. 
Fukuda et al. \cite{fukuda17_interspeech} showed that the accuracy of knowledge distillation can be improved by ensemble fusion of multiple teacher networks in the Aurora 4 test set (speech recognition task under noise condition). Thus, the methods for utilizing weakly labeled or unlabeled data can be divided into two categories: using knowledge distillation and using pseudo-labeling with tentative model. In this paper, we propose a method based on knowledge distillation. Our method differs from existing methods in that it performs knowledge distillation in the student's network in addition to knowledge distillation between student-teacher networks. Therefore, our method can collaborate with other proposed methods such as \cite{9312148}.

Recently, an online knowledge distillation method \cite{9412615, Guo_2020_CVPR, Asif2020}, an ensemble-based model for knowledge distillation into sub-networks, has been proposed, and it has achieved high accuracy in image recognition tasks. This paper adopts a knowledge distillation model based on semi-supervised learning called ``peer collaborative learning'' (PCL) \cite{Wu_Gong_2021}. Originally, PCL was applied to image recognition tasks, and its effectiveness has been shown. In this study, we apply it to the PSED task for the first time. The main feature of this PCL model is that it combines the advantages of both the online knowledge distillation method and the mean-teacher method. In other words, in the framework of the student-teacher model, the stronger teacher model that accumulate knowledge of the student model stabilizes training of the student model, and the student model itself is equipped with a more accurate sound event detection function by the knowledge distillation using ensemble learning with multiple branching sub-networks inside the student model. This paper also proposes an original method to design the sub-networks inside the student model depending on data augmentation methods for input sound data. 

In the evaluation experiments, we used the test set used in the DCASE 2019 Task 4. As a baseline method, we use the mean-teacher method provided by the Task 4 organizer \cite{Turpault2019}. The online knowledge distillation method \cite{9412615} is used as a comparison method with the proposed method. As a result of the experiment, we obtained a significant improvement in the PCL approach over the baseline, with around 10\% improvement in F1-score, which is a public evaluation index for PSED task. The PCL also showed an accuracy improvement of about 1\% compared to the online knowledge distillation method. We also found that the suitable design of the sub-networks based on the data augmentation methods within the student model could be improved. 

The contributions of this paper are summarized as follows: 
\vspace*{-1mm}
\begin{itemize}
  \itemsep -1pt
\item We first demonstrate the usefulness of PCL in the PSED task.
\item We also show that the accuracy of the model can be improved by designing the internal sub-networks of the student model based on data augmentation methods.
\end{itemize}

\begin{figure}[tb]
  \centering
  \includegraphics[width=0.8\columnwidth]{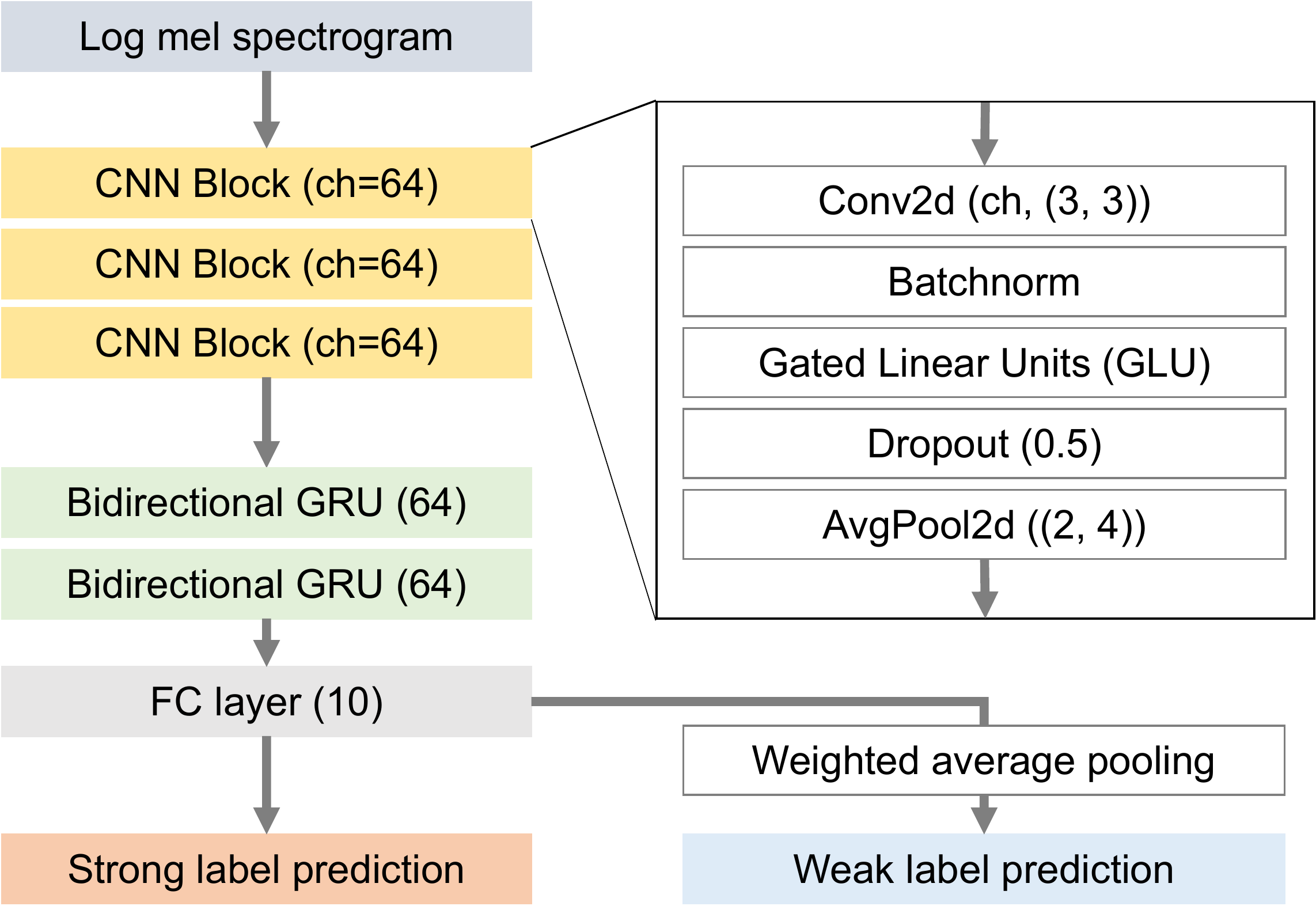}
  \vspace*{-2mm}
  \caption{The model architechture of the student and teacher models. }
  \label{fig02}
\end{figure}

%
%
\section{Sound Event Detection Models}
\label{sec:2}

\subsection{Mean-Teacher Model}

The mean-teacher method \cite{Turpault2019} has been adopted as the baseline method for the DCASE 2019 Task 4. Figure \ref{fig01}, already shown, provides an overview of the mean-teacher method, and Fig. \ref{fig02} shows the specific network structure of the student and teacher models.

This method uses two models with the same structure, and the two models are trained in a consistent way to maintain the consistency of their outputs. First, normal input sound data (sound signal) is input into the student model. On the other hand, the data with noise added to the sound signal is input to the teacher model, and each output result is obtained. Then, for the student model, the model parameters are updated using the classification loss (cross-entropy) for the assigned labels and the consistency loss (minimum square error) with the output of the teacher model. Finally, the parameters of the teacher model are updated by an exponential moving average of the parameter values of the student model. This exponential moving average is a parameter-copying method that gives weight to the most recently trained weight parameters of the model and considers model parameters that have been trained in the past. This makes the teacher model a temporal ensemble model of the student model, i.e., the teacher model stores the past learning states of the student model.

As this process is repeated, the teacher model is gradually trained into a model that reflects the learning process of the student model and finally becomes a role model for the student model, and can guide the training of the student model through the calculation of consistency loss.
The loss for parameter update based on back propagation are calculated based on total sum of $L_{BCE}$ and $L_{MSE}$, shown in Fig. \ref{fig01}.

\subsection{Online Knowledge Distillation}

\begin{figure*}[tb]
  \centering
  \includegraphics[width=1.7\columnwidth]{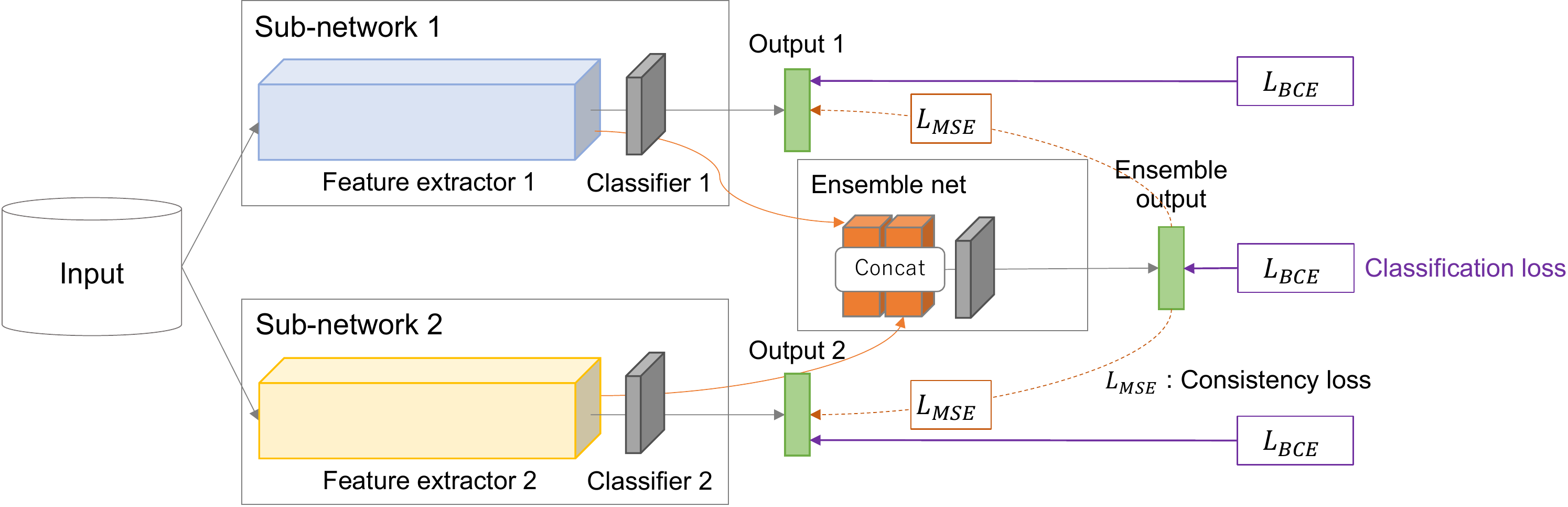}
  \vspace*{-2mm}
  \caption{A framework for training an online knowledge distillation model using ensemble net.}
  \label{fig03}
\end{figure*}

\begin{figure*}[tb]
  \centering
  \includegraphics[width=1.95\columnwidth]{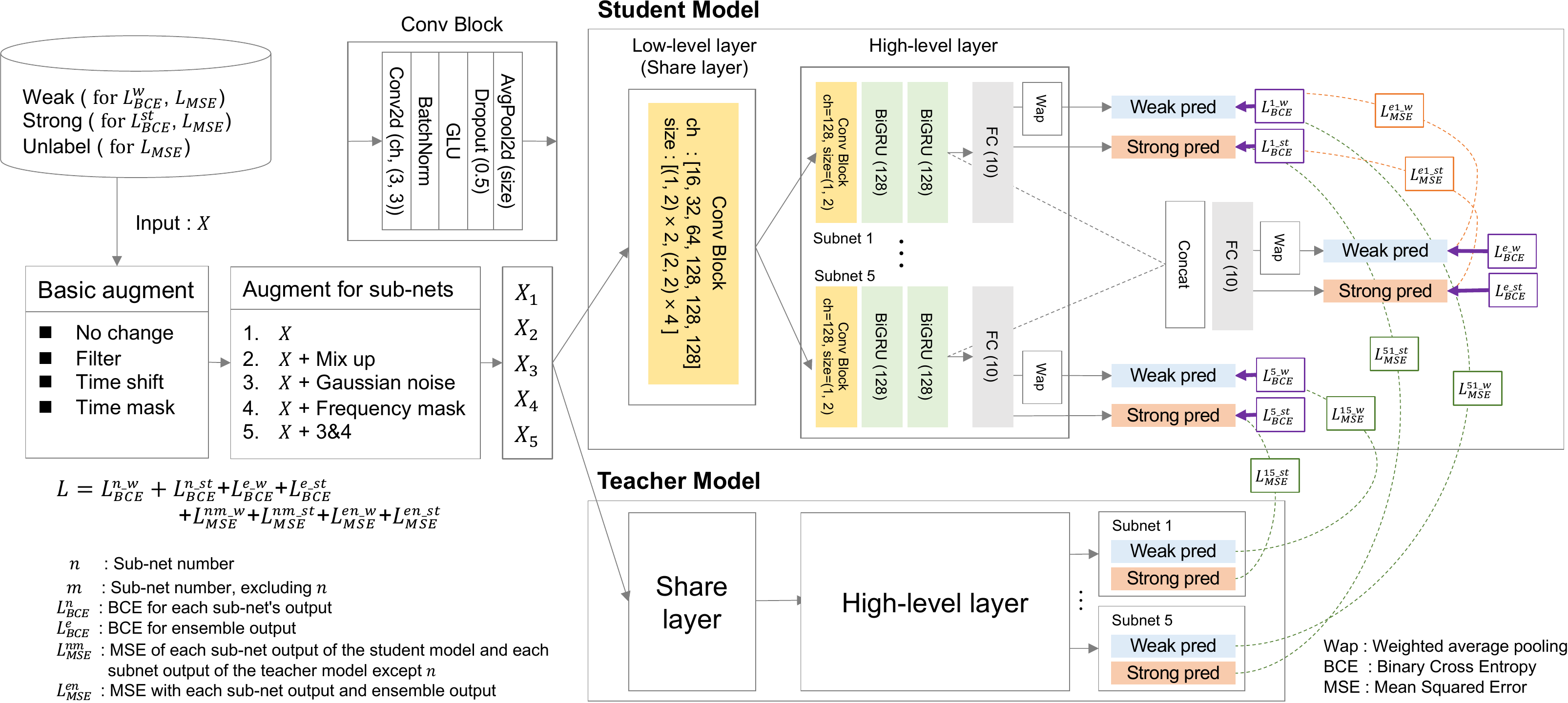}
  \vspace*{-2mm}
  \caption{The framework of peer collaborate learning, the layers that compose each model, and the calculation of losses.}
  \label{fig04}
\end{figure*}

Figure \ref{fig03} shows the training framework of the online knowledge distillation method. Figure \ref{fig03} shows an example of the whole network with two sub-networks. The number of sub-networks can be increased, and the number of sub-networks is set to five in this paper. The detailed layer structure is the same as that of the PCL method in Fig. \ref{fig04}, and the student model of the PCL method and the network structure of the online knowledge distillation model in Fig. \ref{fig03} are consistent.

This method uses parallel sub-networks inside the student model and an ensemble net that combines the feature representations extracted by each sub-network. The ensemble net is a very powerful feature extraction model because it integrates the output of each sub-network. By distilling the output of this ensemble net into each sub-network, they can extract complementary features for sound event classification. This means that the feature extraction performance of each sub-network is improved, and the generalization performance of the entire model is improved. This knowledge distillation using ensemble nets is very effective.

The number of sub-networks will be explained in the next section. For simplicity, we adopt a method that determines the sub-networks according to what kind of data augmentation process has been adopted on the input acoustic data. Note that the loss used to train the knowledge distillation model is the sum of the consistency losses and classification losses of the outputs of the ensemble net and sub-networks.

\subsection{Peer Collaborative Learning}

This section describes the PCL method, which combines the advantages of both the online knowledge distillation method and the average teacher method. The entire network of the model is shown in Fig. \ref{fig04}.

First of all, the input acoustic data is subjected to a data augmentation process. In this study, we prepared mixup, Gaussian noise addition, and frequency mask as data augmentation methods. In practice, the following five processes are employed: 1. no data augmentation, 2. mixup \cite{Zhang2018}, 3. Gaussian noise addition, 4. frequency mask \cite{park19e_interspeech} and 5. Gaussian noise + frequency mask.

In the student model, we adopt a method called the peer ensemble model. The characteristic of this method is that the network is branched depending on the data augmentation process. When inputting data into the model, each data after data augmentation passes through lower layers shared by sub-networks. The features of each data extracted in the lower shared layers are then input to the respective sub-network corresponding to the training of that data. After each sub-network extracts the features of the input data, it passes through the classifier layer (fully connected layer) and outputs the sound event prediction results. Here, as in Section 2.2 for online knowledge distillation, ensemble feature extraction is achieved by combining the features finally extracted by each sub-network. The classifier using these ensemble features is treated as an ensemble net and outputs the acoustic event prediction results. This ensemble output is then distilled into the output of each sub-network. In addition, the parameters of the teacher model are also updated using the method described in Section 2.1. Note that ensemble nets are not used in the teacher model.

The point of this method is to increase the generalizability of the acoustic feature representation extracted by the sub-networks according to the data augmentation methods. In addition, the proposed method incorporates the advantages of traditional online knowledge distillation methods in that the sub-networks can extract complementary acoustic feature representations from each other, because the ensemble net, which consolidates the knowledge of each sub-network, distills the knowledge for the sub-networks inside the student model. 
Furthermore, knowledge distillation is performed between sub-networks to make the training between sub-networks more consistent. The knowledge distilled here is not the knowledge of the sub-networks in the student model, but the knowledge of the more powerful teacher model built by the mean-teacher method, which allows for more stabilized training. 

\section{Experiments}
\subsection{Experimental Setup}
\subsubsection{Dataset and evaluation metrics}

The dataset consists of audio clips with a maximum length of 10 seconds that were recorded in a home environment or synthesized to assume a home environment. The number of classes of sound events is 10. The training dataset consists of three types of datasets: weakly labeled set, unlabeled set, and strongly labeled set. The weakly labeled set and the unlabeled set are sounds obtained from Audio Set \cite{Gemmeke2017audioset}, while the strongly label set is a synthetic sound generated by Scaper \cite{salamon2017scaper}. The weakly labeled, unlabeled, and strongly labeled sets contain 1578, 14412, and 2045 audio clips, respectively. In this paper, two types of datasets are used to evaluate the model: strongly labeled validation set and strongly labeled test set. The validation set is a combination of the validation set and the evaluation set from the DCASE 2018 Task 4 \cite{Serizel2018}, and the test set is part of Audio Set. The validation and test datasets contain 1168 clips and 692 clips, respectively.

The event-based F1-score \cite{Mesaros2016_MDPI} was used as the evaluation measure. This is a measure of how accurately the acoustic event intervals were estimated for the audio clips in the validation and test set. The evaluation program was provided by the Task 4 organizer and we used it\footnote{\url{https://github.com/TUT-ARG/sed_eval}}.

\subsubsection{Models and training condition}

In this paper, we conducted the following six experiments to evaluate the effectiveness of the proposed method. First, as a baseline method, we use the mean-teacher method used in the DCASE 2019 Task4, which was described in Section 2.1. In the DCASE 2020 Task 4, a slightly improved baseline model has been proposed \cite{Turpault2020a}, denoted as the baseline-advanced (adv). Next, as a comparison method, we use the online knowledge distillation method (Oneline KD) described in Section 2.2, which is a knowledge distillation method using sub-networks and ensemble net. The number of sub-networks was set to five to accommodate the data augmentation process. For the PCL method, three models were prepared: five sub-networks with data augmentation (PCL w/ DA), five sub-networks without data augmentation (PCL w/o DA), and PCL w/ DA without ensemble net (PCL w/o emsemble). In all the models, basic data augmentation processes such as frequency filter, time shift, and time mask are applied. Note that Online KD, PCL w/ DA model, and PCL w/o emsemble apply mixup, etc., in addition to these basic data augmentation processes for branching the sub-networks in the student model.

The baseline model structure is shown in Fig. \ref{fig02}. As input acoustic features, a 64-dimensional log Mel spectrogram was extracted from an audio clip recorded at 44.1 kHz using a window function with a width of 2048 points and 511 point hops. The optimization function used to train the model was Adam, and the learning rate was set to 0.001.

The baseline-adv. model is an improved version of the baseline model (see \cite{Turpault2020a} for details of the model), and the input features are different from the baseline. The model structures used in the Oneline KD and PCL methods are shown in Fig. \ref{fig03} and \ref{fig04}, respectively. The input acoustic features for these models, including the baseline-adv., are 128-dimensional log Mel spectrograms computed from audio clips downsampled to 16 kHz and cut with a width of 2048 points and 255 point hops window function. The optimization function was Adam, and the learning rate follows a ramp-up strategy \cite{LaineA17}, where the maximum learning rate was set to reach 0.001 after 50 epochs. Note that we used the dataset of DCASE 2019 Task 4, so the experimental conditions of DCASE 2019 are as close as possible to those of DCASE 2020 to emulate the experimental environment.

\vspace*{-2mm}
\subsection{Results}

Table \ref{tab:result} shows the experimental results for the two test sets. 

First, it can be seen that the proposed method, PCL w/ DA, has the highest performance. This demonstrates the usefulness of the proposed method. Second, we can also see that both the Online KD and PCL w/o emsemble models also show significant improvement compared to the two baselines. This can be attributed to the significant effect of knowledge distillation by the ensemble net and also knowledge distillation from the teacher model to the student model in the mean-teacher framework. Besides, when comparing Online KD, PCL w/o emsemble, and PCL w/ DA, the F1-score was improved by 1 to 2\% compared to the case where each distillation method was separately applied. Therefore, it is clear that the fusion of each knowledge distillation method is more effective in improving the model than treating each method independently.

Furthermore, it is evident from the comparison between PCL w/ DA and PCL w/o DA that the model is improved by building sub-networks that depend on the data augmentation process.

These experimental results indicate two things: the PCL method is an effective method for the PSED task, and the accuracy of the model can be improved by designing the internal network of the student model based on the data augmentation method for acoustic data.

\begin{table}
  \centering
  \caption{Sound event detection performance for each model (F1-score [\%]). The numbers in parentheses are the published values on the DCASE website}
  \label{tab:result}
  \vspace*{1mm}
  \begin{tabular}{l|cc} \hline\hline
  Model & Validation set & Test set \\ \hline
  Baseline & 25.9 (23.7) & 31.1 (29.0) \\
  Baseline-adv. & 34.7 (34.8) & 36.2 \\
  Online KD & 43.1 & 43.4 \\
  PCL w/ DA (\textit{proposed}) & {\bf 43.8} & {\bf 44.2} \\
  PCL w/o DA & 41.7 & 42.4 \\
  PCL w/o emsemble & 41.9 & 41.0 \\ \hline\hline
  \end{tabular}
\end{table}

\section{Conclusions}

In this paper, we proposed the knowledge distillation method, PCL, for the PSED task, which makes effective use of weakly labeled and unlabeled data.

The PCL method is different from previous knowledge distillation methods for PSED tasks in that it uses multiple sub-networks and an ensemble net that combine them in the student model. As a result of experiments using the test set of DCASE 2019 Task 4, we confirmed the effectiveness of knowledge distillation based on ensemble net.  In addition, the PCL approach, incorporating the existing the mean-teacher method, further improved the performance of PSED. We also found that the model with better generalization performance can be trained by changing the input sub-network according to the data augmentation method of the input sound data.

In future work, we are going to implement and experiment with new knowledge distillation methods, such as collaboration with other knowledge distillation methods (e.g. \cite{9312148}).

\ninept
\bibliographystyle{IEEEbib}
\bibliography{paper}

\begin{thebibliography}{10}

\bibitem{DCASE}
``Detection and classification of acoustic scenes and events,''
  \url{http://dcase.community/},
\newblock [Online; accessed 7-Oct-2021].

\bibitem{Turpault2019}
Nicolas Turpault, Romain Serizel, Justin Salamon, and Ankit~Parag Shah,
\newblock ``Sound event detection in domestic environments with weakly labeled
  data and soundscape synthesis,''
\newblock in {\em Proc. of the Detection and Classification of Acoustic Scenes
  and Events 2019 Workshop (DCASE2019)}, October 2019, pp. 253--257.

\bibitem{DCASE2019Workshop}
Michael Mandel, Justin Salamon, and Daniel P.~W. Ellis,
\newblock {\em Proceedings of the Detection and Classification of Acoustic
  Scenes and Events 2019 Workshop (DCASE2019)},
\newblock October 2019.

\bibitem{DCASE2020Workshop}
Nobutaka Ono, Noboru Harada, Yohei Kawaguchi, Annamaria Mesaros, Keisuke Imoto,
  Yuma Koizumi, , and Tatsuya Komatsu,
\newblock {\em Proceedings of the Fifth Workshop on Detection and
  Classification of Acoustic Scenes and Events (DCASE 2020)},
\newblock Tokyo, Japan, November 2020.

\bibitem{Mesaros2016_MDPI}
Annamaria Mesaros, Toni Heittola, and Tuomas Virtanen,
\newblock ``{Metrics for Polyphonic Sound Event Detection},''
\newblock {\em Applied Sciences}, vol. 6, no. 6, pp. 162, 2016.

\bibitem{44873}
Geoffrey Hinton, Oriol Vinyals, and Jeffrey Dean,
\newblock ``{Distilling the Knowledge in a Neural Network},''
\newblock in {\em reprint arXiv:1502.02531}, 2015.

\bibitem{8639635}
Vimal Manohar, Pegah Ghahremani, Daniel Povey, and Sanjeev Khudanpur,
\newblock ``{A Teacher-Student Learning Approach for Unsupervised Domain
  Adaptation of Sequence-Trained ASR Models},''
\newblock in {\em Proc. of 2018 IEEE Spoken Language Technology Workshop},
  2018, pp. 250--257.

\bibitem{Lin2019}
Liwei Lin, Xiangdong Wang, Hong Liu, and Yueliang Qian,
\newblock ``{Guided Learning Convolution System for DCASE 2019 Task 4},''
\newblock in {\em Proc. of the Detection and Classification of Acoustic Scenes
  and Events 2019 Workshop (DCASE2019)}, October 2019, pp. 134--138.

\bibitem{Park2019b}
Hyoungwoo Park, Sungrack Yun, Jungyun Eum, Janghoon Cho, and Kyuwoong Hwang,
\newblock ``{Weakly Labeled Sound Event Detection using Tri-training and
  Adversarial Learning},''
\newblock in {\em Proc. of the Detection and Classification of Acoustic Scenes
  and Events 2019 Workshop (DCASE2019)}, October 2019, pp. 184--188.

\bibitem{Ebbers2020}
Janek Ebbers and Reinhold Haeb-Umbach,
\newblock ``{Forward-Backward Convolutional Recurrent Neural Networks and
  Tag-Conditioned Convolutional Neural Networks for Weakly Labeled
  Semi-Supervised Sound Event Detection},''
\newblock in {\em Proc. of the Detection and Classification of Acoustic Scenes
  and Events 2020 Workshop (DCASE2020)}, November 2020, pp. 41--45.

\bibitem{9312148}
Nam~Kyun Kim and Hong~Kook Kim,
\newblock ``{Polyphonic Sound Event Detection Based on Residual Convolutional
  Recurrent Neural Network With Semi-Supervised Loss Function},''
\newblock {\em IEEE Access}, vol. 9, pp. 7564--7575, 2021.

\bibitem{fukuda17_interspeech}
Takashi Fukuda, Masayuki Suzuki, Gakuto Kurata, Samuel Thomas, Jia Cui, and
  Bhuvana Ramabhadran,
\newblock ``{Efficient Knowledge Distillation from an Ensemble of Teachers},''
\newblock in {\em Proc. Interspeech 2017}, 2017, pp. 3697--3701.

\bibitem{9412615}
Jangho Kim, Minsung Hyun, Inseop Chung, and Nojun Kwak,
\newblock ``{Feature Fusion for Online Mutual Knowledge Distillation},''
\newblock in {\em Proc. of ICPR 2021}, 2021, pp. 4619--4625.

\bibitem{Guo_2020_CVPR}
Qiushan Guo, Xinjiang Wang, Yichao Wu, Zhipeng Yu, Ding Liang, Xiaolin Hu, and
  Ping Luo,
\newblock ``{Online Knowledge Distillation via Collaborative Learning},''
\newblock in {\em Proc. of CVPR 2020}, June 2020, pp. 11020--11029.

\bibitem{Asif2020}
Umar Asif, Jianbin Tang, and Stefan Harrer,
\newblock ``{Ensemble Knowledge Distillation for Learning Improved and
  Efficient Networks},''
\newblock in {\em Proc. of 24th European Conference on Artificial
  Intelligence}, 2020, pp. 1--8.

\bibitem{Wu_Gong_2021}
Guile Wu and Shaogang Gong,
\newblock ``Peer collaborative learning for online knowledge distillation,''
\newblock {\em Proc. of AAAI 2021}, vol. 35, no. 12, pp. 10302--10310, May
  2021.

\bibitem{Zhang2018}
Hongyi Zhang, Moustapha Cisse, Yann~N. Dauphin, and David Lopez-Paz,
\newblock ``{mixup: Beyond Empirical Risk Minimization},''
\newblock in {\em Proc. of ICLR 2018}, 2018, pp. 1--13.

\bibitem{park19e_interspeech}
Daniel~S. Park, William Chan, Yu~Zhang, Chung-Cheng Chiu, Barret Zoph, Ekin~D.
  Cubuk, and Quoc~V. Le,
\newblock ``{SpecAugment: A Simple Data Augmentation Method for Automatic
  Speech Recognition},''
\newblock in {\em Proc. of INTERSPEECH 2019}, 2019, pp. 2613--2617.

\bibitem{Gemmeke2017audioset}
Jort~F. Gemmeke, Daniel P.~W. Ellis, Dylan Freedman, Aren Jansen, Wade
  Lawrence, R.~Channing Moore, Manoj Plakal, and Marvin Ritter,
\newblock ``{Audio Set: An ontology and human-labeled dataset for audio
  events},''
\newblock in {\em Proc. of ICASSP 2017}, 2017, pp. 776--780.

\bibitem{salamon2017scaper}
Justin Salamon, Duncan MacConnell, Mark Cartwright, Peter Li, and Juan~Pablo
  Bello,
\newblock ``{Scaper: A library for soundscape synthesis and augmentation},''
\newblock in {\em 2017 IEEE Workshop on Applications of Signal Processing to
  Audio and Acoustics (WASPAA)}. IEEE, 2017, pp. 344--348.

\bibitem{Serizel2018}
Romain Serizel, Nicolas Turpault, Hamid Eghbal-Zadeh, and Ankit~Parag Shah,
\newblock ``{Large-scale weakly labeled semi-supervised sound event detection
  in domestic environments},''
\newblock in {\em Proc. of the Detection and Classification of Acoustic Scenes
  and Events 2018 Workshop (DCASE2018)}, November 2018, pp. 19--23.

\bibitem{Turpault2020a}
Nicolas Turpault and Romain Serizel,
\newblock ``Training sound event detection on a heterogeneous dataset,''
\newblock in {\em Proceedings of the Detection and Classification of Acoustic
  Scenes and Events 2020 Workshop (DCASE2020)}, November 2020, pp. 200--204.

\bibitem{LaineA17}
Samuli Laine and Timo Aila,
\newblock ``{Temporal Ensembling for Semi-Supervised Learning},''
\newblock in {\em Proc. of {ICLR} 2017}, 2017, pp. 1--13.

\end{thebibliography}

\end{document}